\documentclass[amsmath,amssymb,aps,prl,showpacs,twocolumn,groupedaddress]{revtex4-1}

\usepackage{graphicx}
\usepackage{dcolumn}

\begin{document}

\title{Quenching of Impurity Spins at Cu/CuO Interfaces:
An Antiferromagnetic Proximity Effect}

\author{Ko Munakata}
\email{komun@stanford.edu}
\author{Theodore H. Geballe}
\author{Malcolm R. Beasley}
\affiliation{Department of Applied Physics, Stanford University,
Stanford, California 94305, USA}
\affiliation{Geballe Laboratory for Advanced Materials, Stanford University,
Stanford, California 94305, USA}

\date{\today}

\begin{abstract}
It is observed that the magnetoconductance of bilayer films of copper (Cu)
and copper monoxide (CuO) has distinct features compared of that of Cu films on
conventional band insulator substrates.
We analyze the data above 2 K
by the theory of weak antilocalization in two-dimensional metals and
suggest that spin-flip scatterings by magnetic impurities
inside Cu are suppressed in Cu/CuO samples.
Plausibly the results imply a proximity effect of antiferromagnetism
inside the Cu layer, which can be understood in the framework of
Ruderman-Kittel-Kasuya-Yoshida (RKKY) interactions.
The data below 1 K, which exhibit slow relaxation
reminiscent of spin glass, are consistent with this interpretation.

\end{abstract}


\maketitle
As the technology to synthesize high-quality thin films
and thin film interfaces steadily
improves, there has been an extensive search for novel physical
properties in thin film heterostructures in the condensed matter physics community.
In fact, numerous heterostructure interfaces have been found to exhibit
unique phenomena that are not present in bulk materials.
Some prominent examples include
the exchange bias effects in antiferromagnet/ferromagnet interfaces
\cite{QTM},
high-mobility two-dimensional electron gases in
semiconductor and complex oxide heterostructures \cite{HaPRB78,OhNATURE04},
and various proximity effects.
The proximity effect at solid state interfaces
can be defined as a mutual induction of certain physical properties
from one material into an adjacent one across their interface.
The most famous example is that of superconductivity,
where superconducting pairs
are induced in a neighboring normal metal while
normal electrons in the metal permeate the superconductor \cite{McPR68, BuRMP05}.

At the interfaces between a metal and a non-superconducting material, especially
an insulator, one might naively expect no proximity effect besides
a simple transfer of charges and development of a Schottky barrier.
In this paper, however, we present evidence for a new proximity effect
that arises between a normal metal
and an antiferromagnetic (AF) charge-transfer insulator.
Specifically, we show evidence for the creation of AF
spin ordering in a normal metal due to the proximity effect
through spin-spin interactions with an
AF charge transfer insulator.
The existence of such a proximity effect has been anticipated theoretically
\cite{TheoryAPE}.
The heterostructure of a copper (Cu) thin film and
a copper monoxide (CuO) thin film was synthesized
as a potential model system for such a proximity effect \cite{CuprateInterface}.
This Cu/CuO bilayer exhibits distinct features in magnetotransport
compared to a Cu thin film on a conventional band insulator substrate.
The magnetoconductance of both films above 2 K can be analyzed by
the theory of weak antilocalization
and indicates the quenching of spin-flip scatterings
by magnetic impurities inside the Cu in proximity to CuO.
This non-local effect in magnetotransport by an AF insulator can be naturally
interpreted as a consequence of AF spin ordering induced in the Cu. 

Smooth thin Cu/CuO bilayer films were synthesized on magnesium oxide (MgO) substrates.
We first cleaned a MgO (001) substrate and deposited a 21 nm CuO film
as described in the supplementary material \cite{supplementary}.
After the deposition of CuO,
the sample was cooled down to room temperatures under plasma-excited
atomic oxygen flux.
We then turned off the atomic oxygen flux and deposited a 3 nm Cu film
in vacuum by electron beam evaporation.
As a comparison, we also synthesized 3 nm Cu films using the same
Cu source on several different band insulator substrates (MgO, Al$_2$O$_3$, Si),
which we collectively call Cu/BI films because all the films
behaved in the similar way in the transport measurements.
The transport properties
were measured with
a Quantum Design Physical Property Measurement System.

\begin{figure}
\includegraphics[width=3.375in]{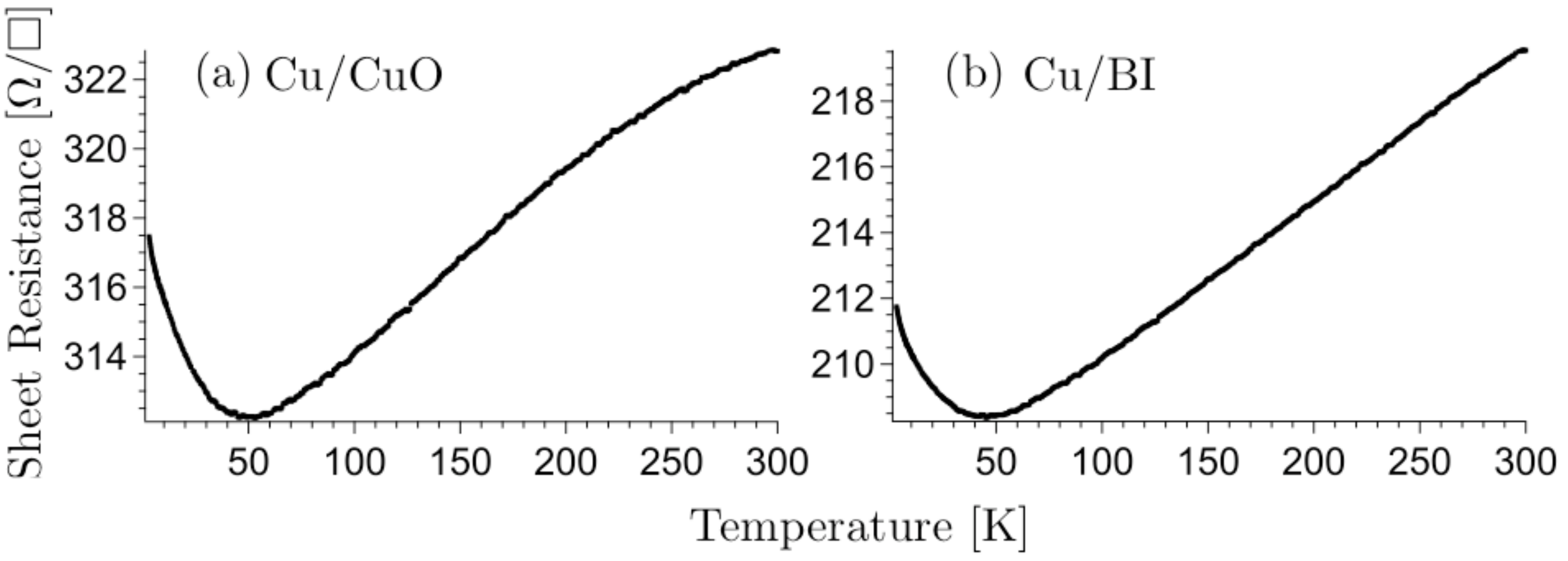}
\caption{\label{fig:Sh} Temperature dependence of sheet resistance
of (a) the Cu/CuO and (b) the Cu/MgO film.
The deviation from the straight line above 200 K in (a) is
due to the resistance of CuO.}
\end{figure}

\begin{figure}
\includegraphics[width=3.375in]{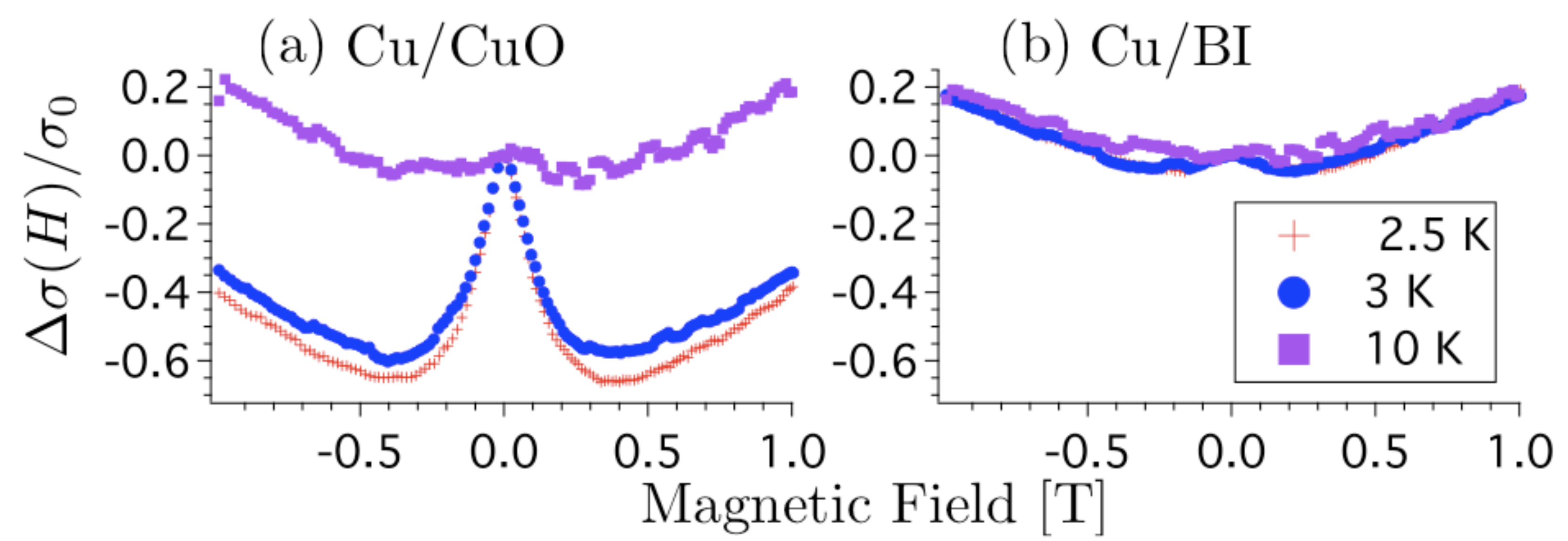}
\caption{\label{fig:MC}
Normalized magnetoconductance at 2.5 K, 3 K, and 10 K of
(a) the Cu/CuO and (b) the Cu/MgO film.
$\Delta\sigma(H)$ is the difference in sheet conductance
between the value under a magnetic field $H$ and the zero-field value,
while $\sigma_0 \equiv e^2/\pi h \simeq 1.23\times 10^{-5} $ S is a constant
for the normalization.
}
\end{figure}

Figs. \ref{fig:Sh}
(a) and (b) represent the sheet resistance of Cu/CuO and
Cu/MgO films, respectively, as functions of temperature.
While the resistance of both films above 50 K increases with temperature
as expected for a simple metal,
both samples have a minimum in the sheet resistance about 50 K.
In order to further examine the transport properties of the two films,
we show in Figs. \ref{fig:MC} (a) and (b) the sheet conductance
as a function of external magnetic
field $H$ perpendicular to the films at different temperatures between 2 K and 10 K.
In both samples, the magnetoconductance is negative
at fields lower than $\sim 0.4$ T and positive
at higher fields, which become more evident at lower temperatures.

These transport properties in two-dimensional metallic systems like Cu
are well known and were extensively examined since the late
1970s and attributed to a weak localization/antilocalization effect
\cite{MaJPSJ81, AbPRB83, BePR84, BeZPB82, FePRB86, supplementary}.
In fact, since CuO and band insulators
have much higher resistance at these temperatures than Cu,
the current must predominantly flow inside the Cu.
This consideration, together with the fact that
both Cu films were deposited from the identical, fully melted
Cu source using the same
e-beam system,
reasonably suggests that there should be
no big difference between the transport properties
of Cu/CuO and Cu/BI films.
However, Fig. \ref{fig:MC} also shows a large difference
in detailed shapes of the curves:
The negative component of the magnetoconductance
at magnetic fields lower than 0.4 T is much more prominent
in the Cu/CuO film than in the Cu/MgO film.
This is the essential experimental finding in this paper.

In order to examine the origin of this difference, we analyze the
data by fitting the magnetoconductance curves to the theoretical equation
for the weak antilocalization effect
\cite{MaJPSJ81, BePR84, Zeeman1}:
\begin{eqnarray*}
\frac{\Delta\sigma(H)}{\sigma_0}  = 
- \frac{3}{2} \bigg\lbrace \ln\frac{\frac{4}{3} H_1 + H_2}{H}
- \psi\Big(\frac{1}{2} + \frac{\frac{4}{3} H_1 + H_2}{H}\Big)&\bigg\rbrace& \\
+  \frac{1}{2}\bigg\lbrace \ln\frac{H_2}{H}
- \psi \Big(\frac{1}{2} + \frac{H_2}{H}\Big) &\bigg\rbrace& \\
\end{eqnarray*}
In the equation above, $\Delta\sigma(H)$ is the difference in sheet conductance
between the value under a magnetic field $H$ and the zero-field value.
There are two fitting parameters
$H_1 \equiv H_{so} - H_s$ and
$H_2 \equiv H_i + 2H_s$,
where $H_i \equiv \hbar/4eD\tau_i$ is the effective field proportional to the
inelastic scattering rate $1/\tau_i$,
$H_s \equiv \hbar/4eD\tau_s$ is proportional to the spin-flip scattering rate
$1/\tau_s$, and
$H_{so} \equiv \hbar/4eD\tau_{so}$ is proportional to the spin-orbit scattering
rate $1/\tau_{so}$.
$\sigma_0 \equiv e^2/\pi h \simeq 1.23\times 10^{-5} $ S is a constant with
the unit of conductance, $D$ is the diffusion
constant for electron motion inside Cu films, and $\psi$ is the digamma function.
We emphasize that this formula and its relatives
have been successfully applied to
many metallic thin films
\cite{AbPRB83, BePR84}
as well as two-dimensional electron gas systems \cite{KoPRL02, CaPRL10},
which supports the reliability of our analysis.

\begin{figure}
\includegraphics[width=3.375in]{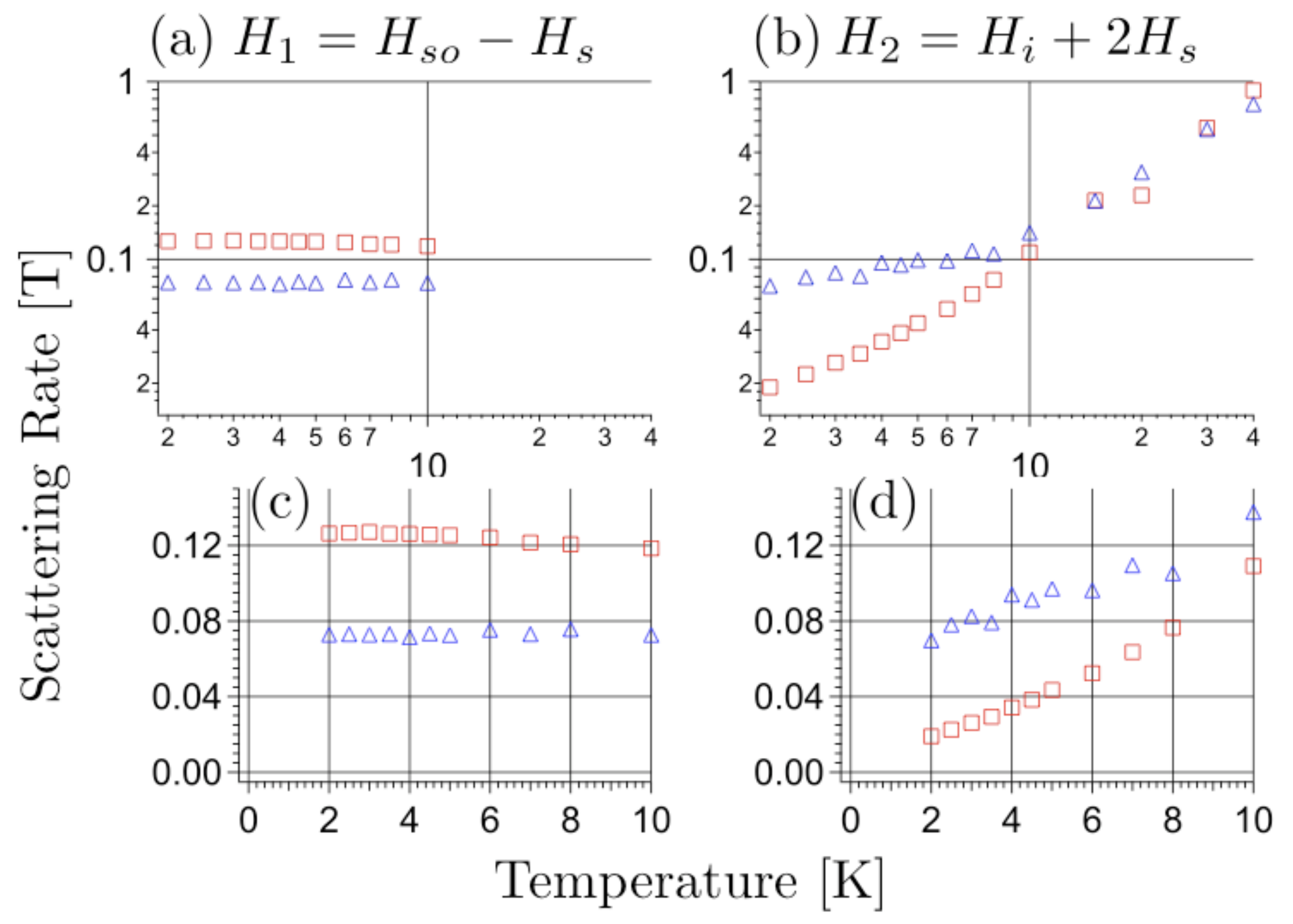}
\caption{\label{fig:BT}
Temperature dependence of (a) $H_1$ in a log scale, (b) $H_2$
in a log scale, (c) $H_1$ in a linear scale, and
(d) $H_2$ in a linear scale.
In each figure, squares represent the Cu/CuO film, while triangles represent
the Cu/MgO film.
}
\end{figure}

We examine the temperature dependence of the different scattering rates
by fitting the experimental curve at each temperature by the theoretical equation.
Figure \ref{fig:BT} (a)
represents the temperature dependence of $H_1$, which is related
to the spin-orbit and spin-flip scattering rates.
The two films have similar values in $H_1$, which do not seem to have significant
temperature dependence.
Since both the spin-orbit scattering and the spin-flip scattering
are expected to be temperature independent \cite{spinflip},
the experimental results that $H_1$ does not exhibit large
temperature dependence assure the validity of our analysis.

On the other hand, the temperature dependence of $H_2$, as shown in
Fig. \ref{fig:BT} (b),
demonstrates the clear difference in transport properties between
Cu/CuO and Cu/MgO films.
While at temperatures higher than 10 K both films show similar decrease of $H_2$ as temperature decreases,
the decrease of $H_2$ of the Cu/MgO becomes much slower
than that of the Cu/CuO film below 10 K.
In fact, the saturation of the decrease in $H_2$ in thin metallic films
including Cu
has been observed in previous studies by other researchers \cite{AbPRB83, BePR84}.
Since $H_2$ is a weighted sum of inelastic and spin-flip scattering rates,
the saturation has been attributed to the presence of small amount of
magnetic impurities which contributes to the spin-flip scattering.
It is therefore natural to speculate that our Cu films also have magnetic
impurities.
In fact, using the secondary ion mass spectrometry \cite{supplementary},
we observed several trace magnetic impurities
(Cr, Fe, Mn, Ni, and Co) in a
much thicker Cu film deposited from the same Cu source. 
What is unexpected, however, is that, even though we deposited Cu on CuO
from the identical Cu source, we do not see the saturation of the decrease
of $H_2$ in the Cu/CuO film.

\begin{table*}
\caption{\label{table:tau}
Summary of the magnetoconductance analysis on our films.
$d$ and $R_s$ represent the thickness and the minimum
sheet resistance, respectively.
The thickness $d$ is estimated from
$dR_s/dT$ (the slope of Fig. \ref{fig:Sh}) between 150 K and 200 K.
For the discussion on why $d$ is smaller than the nominal thickness (3 nm),
please refer to the supplementary material \cite{supplementary}.
Spin-orbit, inelastic, and spin-flip scattering times are evaluated using the data
in Fig. \ref{fig:BT}.
The error range of each value is
simply estimated by the standard error of the linear regression.
For the calculation of the scattering times, we adopted the following parameters
for Cu: electron mass $= 9.1 \times 10^{-31} $ kg;
fermi velocity $= 1.6 \times 10^6$ m/s;
and carrier density $= 8.5 \times 10^{28}$ m$^{-3}$.}
\begin{ruledtabular}
\begin{tabular}{lccccc}
&
$d$ [nm]&
$R_s \lbrack \Omega /\square \rbrack$ &
$\tau_{so}$[$10^{-12}$s]&
$\tau_i T$ [$10^{-11}$s$\cdot$K]&
$\tau_s$ [$10^{-12}$s]\\
\colrule
Cu/CuO & 1.3 & 312 & 1.4 & 2.7 & (7.1$\pm$0.4)$\times 10^1$\\
Cu/MgO & 1.4 & 208 & 1.3 & 2.7$\pm$1.5 & 4.6$\pm$0.8\\
Cu/Al$_2$O$_3$ & 1.5 & 138 & 2.1$\pm$0.3 & 3.3$\pm$2.8 & 6.1$\pm$1.6\\
\end{tabular}
\end{ruledtabular}
\end{table*}

We further observe that $H_2$ of the Cu/CuO film between 2 K and
4 K is roughly proportional to temperature,
though this is not very conclusive due to the narrow range of the measurements.
Since theoretically the inelastic scattering rate by electron-electron scattering
in disordered metals is also expected to roughly scale as $\sim T^1$
\cite{AlJPC82, discrepancy},
this observation implies that $H_2$ in this sample is
dominated not by the spin-flip scattering but by the inelastic scattering.

By plotting $H_2$ as a function of temperature
in a linear scale (Fig. \ref{fig:BT} (d)) and 
linearly extrapolating each curve down to 0K,
the spin-flip scattering time $\tau_s$ of each sample can be estimated.
We can then use the value of $H_1$ (Fig. \ref{fig:BT} (c))
to obtain the spin-orbit scattering time.
The results of the analysis are summarized in Table \ref{table:tau}
for reference.
Table \ref{table:tau} clearly demonstrates that
$\tau_s$ of the Cu/CuO film is anomalously long
compared to that of the Cu/BI films.
On the other hand, the fact that $\tau_{so}$ of each film agrees well further
confirms the validity of our analysis.
We note that, depending on the thickness of the films,
the spin-orbit scattering times of copper films in the literature
roughly range from $10^{-12}$ to $10^{-11}$ s \cite{BeZPB82, FePRB86},
which is consistent with our results.
Table \ref{table:tau} also shows the results of analysis in a
slightly thicker and less disordered Cu/Al$_2$O$_3$ film for comparison.
Although the data are more noisy,
$\tau_s$ of this film is very similar to that
of the Cu/MgO film.
This observation safely excludes the possibility that the magnetic impurities
originate from a surface of any particular BI substrate.

\begin{figure}
\includegraphics[width=3.375in]{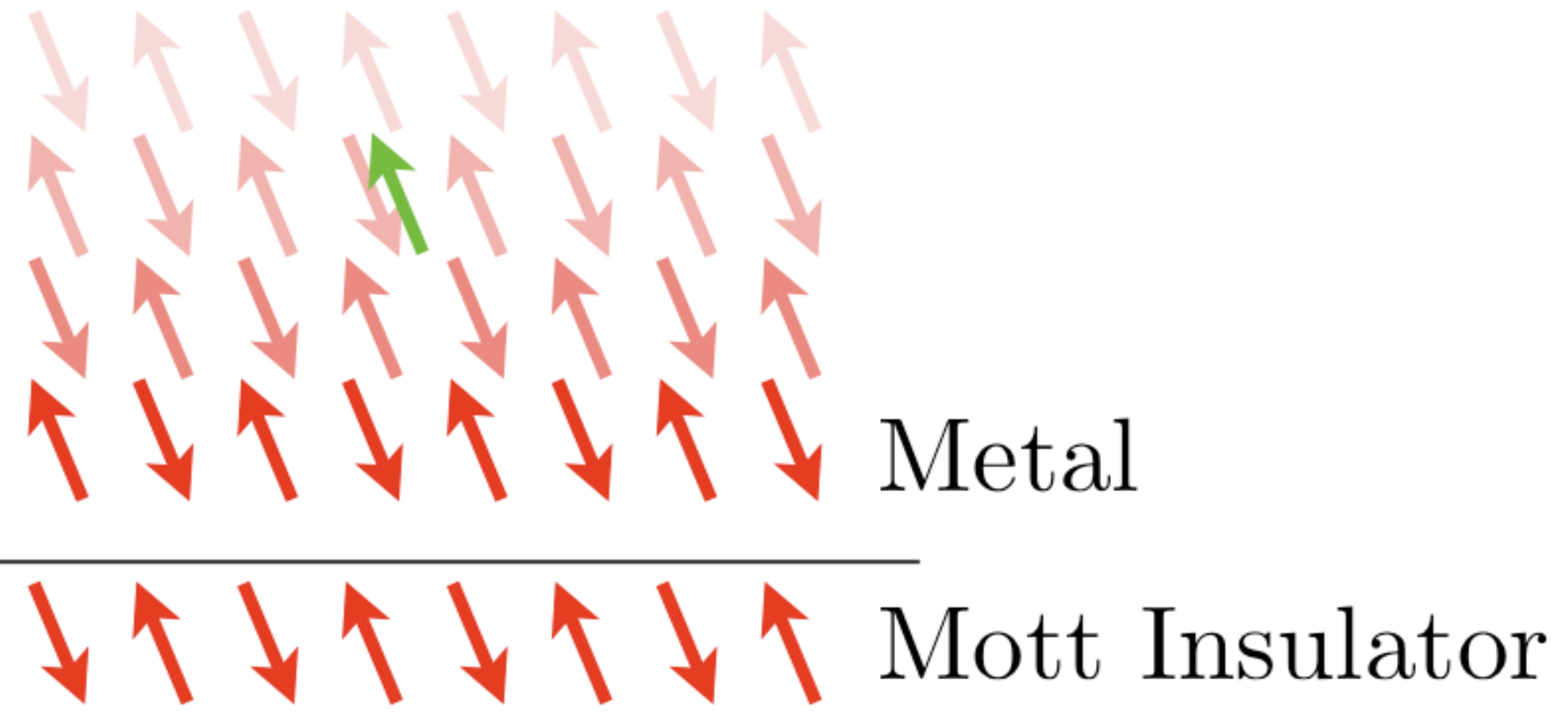}
\caption{\label{fig:APE}
Schematic picture of the proximity effect of antiferromagnetism
implied by our experimental results.
The arrows represent electron spins.
This figure serves as an intuitive understanding of
how the electron spins inside the metal are polarized by the surface
spins of the antiferromagnetic insulator, and
how the spin of each magnetic impurity,
which is represented as a green arrow,
is quenched due to the interactions with neighboring spins
in the metal.
Note that the actual spatial configuration of spin density induced in the Cu
is probably much more complicated than depicted in this figure
due to the low symmetry of the surface spin configuration of the CuO.
}
\end{figure}

All the experimental results presented so far suggest a single story:
while both Cu films contain magnetic impurities, the spin-flip scattering
by the magnetic impurities in the Cu/CuO film
is suppressed due to the adjacent CuO layer.
We can understand this phenomenon in the following way.
Since the spins in CuO are antiferromagnetically aligned below
its N\'{e}el temperature ($\sim 200$ K) as depicted in Fig. \ref{fig:APE},
each nearly-free electron in
the Cu is spin-polarized by the superposition of RKKY interactions
\cite{RuPR54, KaPTP56, YoPR57} from
all the spins on the surface layer of the CuO \cite{afmvector}, which results in an
AF alignment of spins inside the Cu.
In this situation, the spin of each magnetic impurity feels
the spin-polarization of mobile electrons around it through a conventional
exchange interaction.
Such an interaction with polarized spins
naturally creates an energy cost
for the spin-flip process of the magnetic impurity.
When the temperature is lower than this energy cost,
the spin-flip scattering by the magnetic impurity
is exponentially suppressed.

The two copper films measured in a dilution refrigerator
exhibit another characteristic feature below about 1 K,
a hysteresis of magnetoconductance.
While the data are presented elsewhere,
we here note that the presence of the hysteresis implies
the extended relaxation time
in spin glass \cite{QTM} and is probably due to RKKY interactions between
magnetic impurity spins inside the Cu films \cite{SGNote}.
We emphasize that the two copper films with different substrates
have similar magnitudes of hysteresis.
This observation suggests
that the type of magnetic impurities and their concentration
are similar in both films, and is consistent with our interpretation
of the results above 2 K explained above.

In conclusion, through the magnetotransport study above 2 K,
spin-flip scattering is found to be suppressed in Cu/CuO films,
whereas the results in Cu/BI films clearly indicate the presence
of magnetic impurities.
We propose that the observations are indirect evidence
of the proximity effect of antiferromagnetism in the metal.

Even though the effect presented in this paper is subtle, it
might find some interesting applications in the future.
For example, spintronics utilizes the electron's spin and magnetic moment
to affect electrical transport. A common problem in this context is
the undesired relaxation of spin-polarized carriers, possibly due to
magnetic impurities.
Therefore, our experimental
results could have implications for future spintronic experiments.

We thank
A. Tsukada, R.B. Hammond, L. Zhang, N. Breznay and C. Hitzman for
valuable experimental help, and
D. Scalapino and S.A. Kivelson for helpful conversations about the meaning of our results.
This work is supported by the AFOSR and was supported initially by the U.S. DOE.
K.M. is supported by a
Stanford Graduate Fellowship.


\begin{thebibliography}{31}
\bibitem [{The()}]{TheoryAPE}%
  \BibitemOpen
  \href@noop {} {}\bibinfo {note} {Recently there have been explicit
  theoretical calculations of the proximity effect of antiferromagnetism in a
  metal/AF Mott insulator bilayer \cite{ShIJMPB10} and in related systems
  realized by ultracold atoms \cite{SnNJP08}, though we describe the similar
  physics from the point of view of RKKY interactions in Fig.
  \ref{fig:APE}.}\BibitemShut {Stop}%
\bibitem [{Cup()}]{CuprateInterface}%
  \BibitemOpen
  \href@noop {} {}\bibinfo {note} {Our model system is reminiscent of the
  heterostructures of underdoped and overdoped cuprates, which have recently
  been demonstrated to have an enhanced superconducting transition temperature
  \cite{GoNATURE08, YuPRL08}. We speculate that such enhancement of transition
  temperatures, as originally proposed theoretically \cite{EmPRB97, BePRB08},
  might involve a physics similar to the phenomenon presented in this letter if
  the superconductivity in cuprates is due to antiferromagnetic spin
  fluctuations \cite{EmPRB97}.}\BibitemShut {Stop}%
\bibitem [{Zee()}]{Zeeman1}%
  \BibitemOpen
  \href@noop {} {}\bibinfo {note} {Although the original paper by Maekawa and
  Fukuyama \cite{MaJPSJ81} includes the Zeeman splitting term $\gamma$ that is
  proportional to the electron g-factor and external magnetic field $H$, the
  magnetic field we use in our experiments is small enough to ignore this
  term.}\BibitemShut {Stop}%
\bibitem [{spi()}]{spinflip}%
  \BibitemOpen
  \href@noop {} {}\bibinfo {note} {This is only true up to the first order
  scatterings \cite{ASSP}. We assume that the second and higher order spin-flip
  processes are negligible.}\BibitemShut {Stop}%
\bibitem [{dis()}]{discrepancy}%
  \BibitemOpen
  \href@noop {} {}\bibinfo {note} {The measured proportionality constant, which
  according to the theory \cite{AlJPC82} depends on sheet resistance, is
  actually larger by a factor of 7 than what the theory predicts. Although we
  do not understand the reason of this disagreement, similar discrepancy
  between theory and experiment has been reported in the literature
  \cite{GeADP99}. We also note that, if we calculate the proportionality
  constant at 4.5 K for the Cu film of \cite{BeZPB82} and scale it with the sheet
  resistance, the result agrees with our value.}\BibitemShut {Stop}%
\bibitem [{afm()}]{afmvector}%
  \BibitemOpen
  \href@noop {} {}\bibinfo {note} {Since the antiferromagnetic propagation vector
  of CuO is along the (10$\bar{1}$) direction \cite{AiJPC92},
  we assume that the spins in the (111) surface of our film
  do not order ferromagnetically.}\BibitemShut {Stop}%
\bibitem [{SGN()}]{SGNote}%
  \BibitemOpen
  \href@noop {} {}\bibinfo {note} {Strictly speaking, this may not be genuine
  spin glass in the sense that the relaxation time is infinite, because the spin
  glass should not occur in a two-dimensional system from a theoretical point of
  view. Even so, the relaxation time can be long enough to exhibit a hysteresis
  in our experimental time scale.}\BibitemShut {Stop}%
\end{thebibliography}

\providecommand{\noopsort}[1]{}\providecommand{\singleletter}[1]{#1}%
%



\section{Synthesis and Characterizations of Films}

Our copper monoxide (CuO) film was synthesized on a magnesium oxide (001)
substrate by electron-beam evaporation
in a vacuum chamber with the base pressure at least better than
$1\times 10^{-8}$ torr.
In order to prepare smooth and clean substrate before deposition,
an ultrasonically cleaned MgO (001) substrate was annealed first
at $500^{\circ}$C in vacuum for a few hours and further
at $750^{\circ}$C under RF-excited atomic oxygen flux \cite{InAPL99} for
10 minutes.
A few nanometers of homoepitaxial MgO
was then deposited by pulsed laser deposition
using a Mg target at $750^{\circ}$C under the atomic oxygen flux,
which yielded a very smooth and chemically clean MgO surface.
Following the cleaning procedure, the substrate was cooled down to
$500^{\circ}$C, where Cu was deposited using the
electron beam evaporation under the atomic oxygen to synthesize a 21 nm CuO film.

X-ray diffraction study has shown that
the (111) direction of our CuO film is aligned parallel to the
(001) direction of the MgO substrate, in the similar manner to
the works in the literature \cite{WaTSF94}.
Although the CuO film is not single crystalline due to the twinning of CuO
with respect to MgO, the (111) peak of the x-ray diffraction is sharp with the
FWHM of the rocking curve smaller than 0.2$^\circ$, suggesting a good crystalline
quality.
In addition, the atomic force microscope measurement has shown that
the RMS roughness of the surface is only about 0.5 nm, which makes it possible
to synthesize a ultrathin continuous Cu film on top of the CuO.

We also looked for trace magnetic impurities in a 300 nm-thick Cu film deposited
from the same Cu source by using a secondary ion mass spectrometry (SIMS)
\cite{sims}.
Table \ref{table:sims} shows that Cr, Mn, Fe, Ni, and Cr were found
out of the 7 elements investigated.
The total concentration of these magnetic elements is $\sim$10 ppm.
We note that the measured concentration is accurate only up to a factor of $\sim$2,
due to our rough estimate of sensitivity factors.

\begin{table}
\caption{\label{table:sims}
Summary of the SIMS measurement in ppm.}
\begin{ruledtabular}
\begin{tabular}{ccccccc}
$^{51}$V &
$^{52}$Cr &
$^{55}$Mn &
$^{56}$Fe &
$^{58}$Ni &
$^{59}$Co &
$^{102}$Ru \\
\colrule
0 & 0.1 & 1 & 2 & 2 & 4 & 0 \\
\end{tabular}
\end{ruledtabular}
\end{table}

\section{Interpretations of Transport Data}

When one observes a minimum of sheet resistance in a thin metallic film,
there are at least 3 possible explanations for the minimum: the
weak localization/antilocalization, the Kondo effect \cite{ASSP},
or the electron-electron interaction
in a disordered system \cite{LeRMP85}.
It is therefore sometimes difficult to determine which one is significant.
In our study, however, we believe that our magnetoconductance
analysis focusing only on the weak localization/antilocalization
is valid at least above 2 K from the following reasons.

If the Kondo effect played a major role in our data,
the magnetoconductance would not depend strongly on the direction of
the applied magnetic field.
Moreover, the minimum of resistivity due to the Kondo effect would occur
at the same temperature regardless of the thickness of the film.
Neither turned out to be true for our films.

As for the effect of electron-electron interactions,
it is probable that, as shown by \cite{FePRB86},
the temperature dependence of sheet resistance in Fig. 1 does include
a significant contribution from this effect.
However, it does not affect the results of our analysis using the low-field
magnetoconductance.

Finally, the result in Fig. 3 that shows temperature-independence
of H$_1$ strongly implies that our analysis is valid without
taking either the Kondo effect or the electron-electron interaction
into account.

\section{Origin of Magnetic Impurities}
Even though we found the trace magnetic impurities in our Cu source
by the SIMS measurement,
it may not be a sufficient proof that they are the only source of the spin-flip
scattering.
In particular, it is known \cite{FePRB86} that oxidation of Cu by moisture in air
does occur and can affect the transport properties.
We could therefore imagine that oxygen atoms cause the spin-flip scattering as well.
The latter interpretation is consistent with the observation that
the effective thickness we noted in the Table I is much smaller than
the nominal thickness, probably due to the
oxidation of the surface of the films.

We actually attempted to prevent the oxidation of our Cu samples by
depositing an additional capping layer.
However, the choice of
the best material for the capping layer turned out to be difficult.
The first material we tested was Al$_2$O$_3$, but we found that it reduced the reproducibility of sheet resistance, probably because of the oxidation of Cu
during the growth of the capping layer.
We also deposited Si as the capping layer,
but it was found that Si layer deposited at room
temperatures could easily conduct
current and confuse the interpretation of the data.

We however emphasize that,
since $H_1$ is correctly estimated to be temperature-independent,
our analysis of the magnetoconductance by
utilizing the weak antilocalization formula seems valid, regardless of
the oxidation of our samples.
In addition, our main conclusion is independent of the source of spin-flip
scatterings, because it relies only on the comparison between
the Cu/CuO and the Cu/BI films.

\end{document}